\documentclass[prb,preprint,showpacs,superscriptaddress]{revtex4-1}
\usepackage{graphicx}
\usepackage{textcomp}
\usepackage{color}
\begin{document}

\title{Anisotropic critical currents in FeSe$_{0.5}$Te$_{0.5}$ films and the influence of neutron irradiation.}
\author{M.~Eisterer}
\author{R.~Raunicher}
\author{H.~W.~Weber}
\address{Atominstitut, Vienna University of Technology, Stadionallee 2, 1020 Vienna, Austria}
\author{E.~Bellingeri}
\address{CNR-SPIN, corso Perrone 24, 16152 Genova, Italy}
\author{M.~R.~Cimberle}
\address{CNR-IMEM, via Dodecaneso 33, 16146 Genova Italy}
\author{I.~Pallecchi}
\address{CNR-SPIN, corso Perrone 24, 16152 Genova, Italy}
\author{M.~Putti}
\address{CNR-SPIN, corso Perrone 24, 16152 Genova, Italy}
\address{Dipartimento di Fisica, Universit\`{a} di Genova, Via Dodecaneso 33, 16146 Genova, Italy}
\author{C.~Ferdeghini}
\address{CNR-SPIN, corso Perrone 24, 16152 Genova, Italy}
%\ead{eisterer@ati.ac.at}

\begin{abstract}
We report on measurements of the superconducting properties of FeSe$_{0.5}$Te$_{0.5}$ thin films grown on lanthanum aluminate. The films have high transition temperatures (above 19\,K) and sharp resistive transitions in fields up to 15\,T. The temperature dependence of the upper critical field and the irreversibility lines are steep and anisotropic, as recently reported for single crystals. The critical current densities, assessed by magnetization measurements in a vector VSM, were found to be well above $10^9$\,A\,m$^{-2}$ at low temperatures. In all samples, the critical current as a function of field orientation has a maximum, when the field is oriented parallel to the film surface. The maximum indicates the presence of correlated pinning centers. A minimum occurs in three films, when the field is applied perpendicular to the film plane. In the forth film, instead, a local maximum caused by c-axis correlated pinning centers was found at this orientation. The irradiation of two films with fast neutrons did not change the properties drastically, where a maximum enhancement of the critical current by a factor of two was found.     
\end{abstract}
%\pacs{74.70.Xa, 74.62.En, 74.78.-w, 74.25 Sv, 61.80.Hg, 74.25.Op, 74.25.Wy}
\maketitle
%\newpage
\section{Introduction}
Thin superconducting layers are not only important for electronic applications, but also valuable for exploring the material properties. The 11-family \cite{Hsu08} (FeSe$_{1-x}$Te$_{x}$, FeS$_{1-x}$Te$_{x}$) of the recently discovered Fe based superconductors  \cite{Kam08,Asw10} has the simplest crystallographic structure, which consists only of the iron containing layer being responsible for superconductivity. The compounds of all other families have additional atoms between these layers, which influence the superconducting properties. In addition, most of them, in particular those with high transition temperatures, contain poisonous arsenic in the superconducting layers. This is replaced by the less toxic elements selenium and tellurium in FeSe$_{1-x}$Te$_{x}$, which represents a potential advantage in applications. Little was reported on the critical current density in FeSe$_{1-x}$Te$_{x}$ \cite{Tae09,Liu10} single crystals and, to our knowledge, nothing about this very important parameter in thin films. We will report on the field and temperature dependence of the critical current in four FeSe$_{0.5}$Te$_{0.5}$ films, including its anisotropy, and compare the results to literature data on films of other compounds \cite{Mel10,Han10,Iid10,Tar10,Lee10,Kat10,Iid10b,Moh10} (mostly of the 122 family).  Resistive measurements of the upper critical field and the irreversibility field, which complement existing data obtained on single crystals \cite{Kle10,Lei10,Khi10}, will be presented. The influence of neutron irradiation will be discussed in the context of similar neutron irradiation experiments on other compounds (Sm-1111 \cite{Eis09b,Eis10c}, La-1111 \cite{Kar09}, Ba-122 \cite{Eis09c}).             

\section{Experimental\label{secexp}}
The films were deposited on single crystal lanthanum aluminate (LAO) (\textit{001}) substrates in a ultra high vacuum PLD system using a FeSe$_{0.5}$Te$_{0.5}$ bulk target compound prepared by direct synthesis from high purity materials.  The films were deposited at a residual gas pressure of $5\times 10^{-9}$\,mbar at a deposition temperature of 490\,\textdegree C. The quality of the growth was in-situ monitored by Reflection High Energy Electron Diffraction (RHEED) analysis. The laser beam (KrF, 248\,nm) was focused onto a 2\,mm$^{2}$ spot on the target with a fluency of 2 J\,cm$^{-2}$. The repetition rate and  the target-substrate distance were kept fixed at 3\,Hz and 5\,cm, respectively. More details of the growth conditions are given in Refs.~\onlinecite{Bel09,Bel10}.  
Four films (named A, B, C, and D) were investigated in this study. Their typical dimensions are 2.4\texttimes 5\,mm$^2$, only sample C broke during handling, which reduced its length to 3.1\,mm. The thickness of the superconducting layer is about 200\,nm in samples A, B and C and around 150\,nm in sample D. This estimation is based on the growth rate and the deposition time. 

Four samples were selected on the basis of a similar transition temperature, $T_\mathrm{c}$. The screening resistivity measurements were carried out at CNR-SPIN by a four-probe technique, using ultrasonicly bonded electrical contacts and $T_\mathrm{c}$ was defined as the temperature, where the normal state resistivity dropped to 90\%. Soon after the measurement, the samples were sealed in vacuum and shipped to Vienna for further measurements and processing.

Resistive in-field measurements were made in Vienna up to a maximum field of 15\,T. Current and voltage contacts were made with silver-epoxy. With this technique, the contact resistance was rather high (typically several 100\,$\Omega$) and the contacts tended to detach during thermal cycles. Nevertheless, it was not possible to completely remove them after the measurement, which would be necessary in view of the irradiation process. The organic compounds of the resin would most likely decompose during the irradiation and the emerging gases harm the films; thus, it was decided to irradiate only two samples (A and B, denoted as A$^\mathrm{irrad}$ and B$^\mathrm{irrad}$ after the irradiation) and to measure the other (pristine) samples (C and D) for comparison. Making contacts on irradiated samples turned out to be even more difficult, partly due to the required care in handling radioactive materials, and the following resistive measurements were rather noisy. We could not obtain reliable (resistive) data on sample A$^\mathrm{irrad}$ in perpendicular orientation. The upper critical field, $B_\mathrm{c2}$, and the irreversibility field, $B_\mathrm{irr}$, were defined by 90\% and 10\% resistive criteria, respectively. Since the resistivity was not constant above $T_\mathrm{c}$, its temperature dependence was linearly extrapolated to lower temperatures, as a proper reference for the normal state resistivity.  
 
Magnetization measurements were performed in a commercial vector vibrating sample magnetometer (VSM) with a maximum field of 5\,T. The field sweep rate was chosen as 0.5\,T/min, resulting in an electric field of about 0.07\,$\mu$V\,cm$^{-1}$ at the sample edges, when the film plane is oriented perpendicular to the field. The sample was not only measured in the usual perpendicular configuration, but also rotated and measured in nine steps of 10\textdegree\ until the field was parallel to the surface. The angular dependent critical currents were calculated from both components (parallel and perpendicular to the applied field) of the irreversible magnetic moment using the Bean model. Since the magnetic moment of a thin superconducting layer always points into the direction perpendicular to the film plane, the sample can be oriented with high precision, the error being below 1\textdegree. Two peculiarities of magnetization measurements in oblique magnetic fields have to be considered\cite{Tho10,Hen11}. The currents do not always flow under maximal Lorentz force and the effective electric field changes with sample orientation. The influence of the variable Lorentz force currents can in principle be suppressed by choosing the sample length appropriately \cite{Hen11}, but this was not feasible in the present experiment because samples of the required length (about 1\, cm) cannot be produced. However, the aspect ratio of the samples ensures that predominantly the maximum Lorentz force currents were measured, which we consider as the measured quantity and treat the influence of the variable Lorentz force currents as a systematic error. The upper bound for this (angular dependent) error is acceptable: 21, 17, and 19\% for sample A, B and D. Only in sample C it could amount to 35\% in the limit of diverging variable Lorentz force currents \cite{Hen11}. In reality, the effect is certainly smaller, but leads to an overestimation of the maximum Lorentz force currents, the error increasing systematically with the angle $\theta$ between the film normal and the applied magnetic field. Even for an isotropic superconducting layer an \emph{experimental} $J_\mathrm{c}$-anisotropy, $\gamma_J=\frac{J_\mathrm{c}(90^\circ)}{J_\mathrm{c}(0^\circ)}=\frac{J_\mathrm{c}^\parallel}{J_\mathrm{c}^\perp}$ is expected (resulting from the larger variable Lorentz force currents), and an anisotropy of up to about 1.2 (sample A, B, D and 1.35 for sample C) could be in principle an artifact of the measurement technique.   
The electric field, which is induced in the sample during field ramping, scales as $\cos{\theta}$ leading to a monotonous decrease of $J_\mathrm{c}$ with increasing angle. The decreasing electric field competes with the effect of the decreasing Lorentz force (variable Lorentz force currents) and it is a priori not clear which effect dominates (both are highly nonlinear). This crucially depends on the superconducting properties, namely the dependence of the critical currents on the variable Lorentz force and the current-voltage characteristics, $E\propto J^n$. For a $n$-value of 10 (20) $J_\mathrm{c}$ is reduced by 20\% (compared to the electric field criterion corresponding to 0\textdegree) at 84\textdegree (89\textdegree) and the deviation becomes comparable to the upper limit of the variable Lorentz force contribution only near 90\textdegree.  
Although these effects certainly cause some distortion in the angular dependence of 
$J_\mathrm{c}$, it can be excluded that the effects reported below are only caused by the experiment, because the observed changes of $J_\mathrm{c}$ with $\theta$ are well above the theoretical limits or do not agree with the expected behavior of the     
experimental issues. Note that an excellent agreement with transport measurements was found in coated conductors of optimized geometry \cite{Hen11} and a systematic error caused by a changing electric field is well accepted in SQUID measurements. The electric field is even worse defined in SQUID measurements and depends on the magnetic field \cite{Eis00}, which induces a distortion in the derived field dependence of $J_\mathrm{c}$.    

Neutron irradiation was performed in the central irradiation facility of the TRIGA-Mark-äII reactor in Vienna. Samples A and B were sealed into quartz tubes filled with helium at reduced pressure ($\approx$ 150\,mbar) and exposed to the neutron flux for 7 hours and 19 minutes which corresponds to a nominal fast neutron (E$>$0.1\,MeV) fluence \cite{Web86} of $2\times 10^{21}$\,m$^{-2}$. The high energy neutrons introduce various defects ranging from single displaced atoms to defects of several nm \cite{Eis10c}. They are statistically distributed and uncorrelated.

%%%%%%%%%%%%%%%%%%%%%%%%%%%%%%%%%%%%%%%%%%%%%%%%%%%%%%%%%%%%%%%%%%%%%%%%%%%%%%%%%%%

\section{Results and Discussion\label{SecResDis}}

\begin{figure} \centering \includegraphics[clip,width=0.5\textwidth]{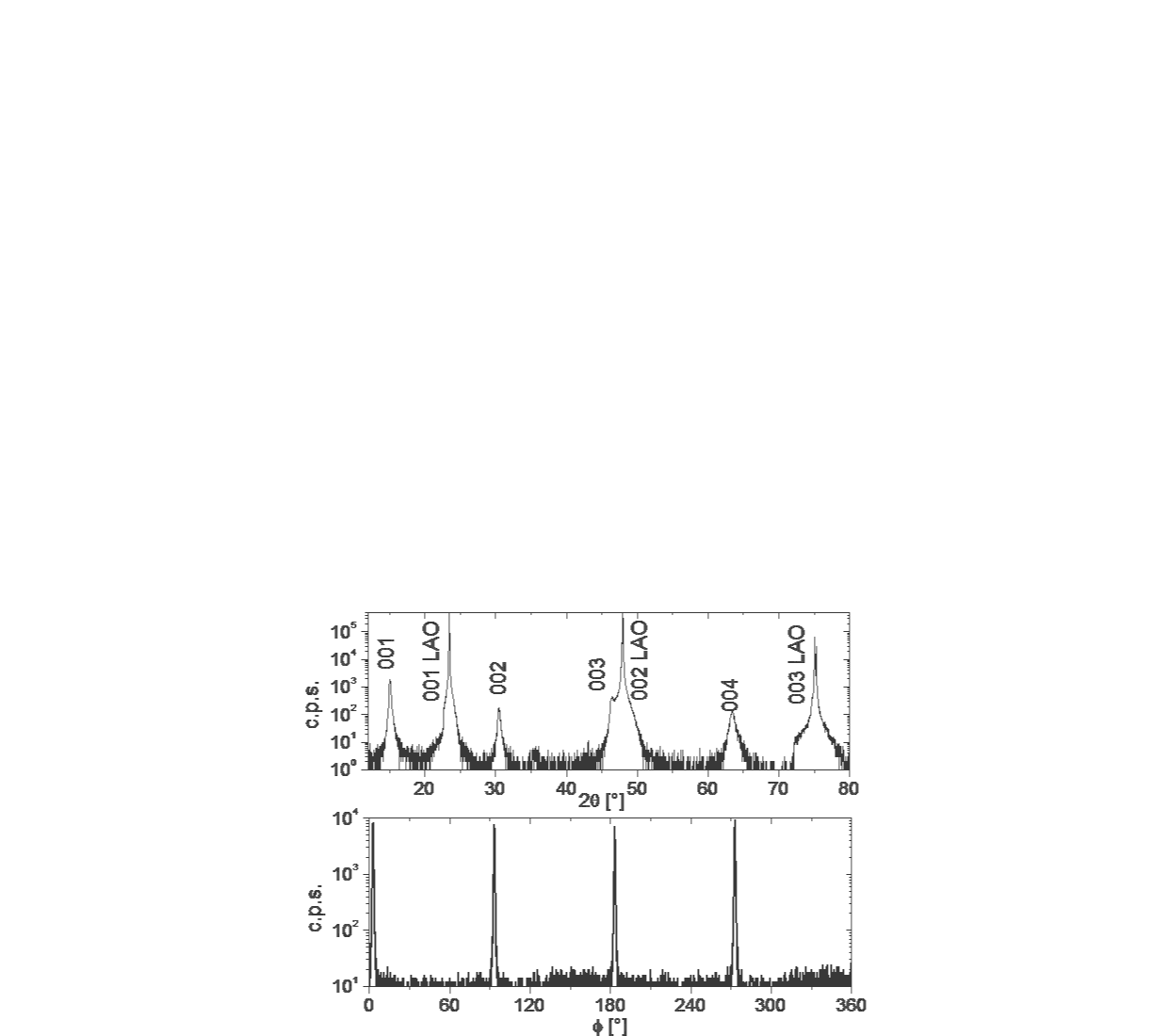}
\caption{Upper panel: XRD  $\theta-2\theta$  scan of a Fe(Se,Te) film deposited on AlLaO$_3$. Only the (\textit{00l}) reflections of the film and substrate are detectable. Lower panel: $\phi$-scan of the (\textit{101}) reflection of the film indicating the epitaxial growth of Fe(Se,Te); the a- and b-axes are found parallel to the substrates axes.} \label{FigXRD}
\end{figure}

In the $\theta-2\theta$ scans, as shown in the upper part of Fig.~\ref{FigXRD}, only the (\textit{00l}) reflections of the film and substrate are present indicating the excellent purity of the phase and the optimal c-axis alignment during the growth. In the lower part of the same figure, $\phi$ scans of the (\textit{101}) reflection of the film (2$\theta=28.13^\circ$; $\chi=33^\circ$) with a FWHM of 0.5\textdegree\ indicate the epitaxial growth of the FeSe$_{0.5}$Te$_{0.5}$ . The  a- and b- axes are parallel to the substrate axes without evidence of any other orientation, thus showing the  high quality of the growth and the full epitaxy with the substrate.

As discussed in Ref. \onlinecite{Bel10}, the FeSe$_{0.5}$Te$_{0.5}$ superconducting phase in these films is under compressive strain partially due to the substrate mismatch but mainly because of the particular growth mode (Volmer-Weber). This compressive strain affects the crystal structure, making the Se,Te tetrahedron closer to the ideal one and leading to a significant increment in the superconducting critical temperature.

The transition temperatures of samples B and D were found to be 19.3\,K, those of samples A and C 19.2\,K and 19.5\,K, respectively.  Since we could not measure the same sample with the same set-up before and after irradiation, we can only estimate the induced change in transition temperature. $T_\mathrm{c}$  was equal (19.3\,K) in samples B and D as derived from the resistivity measurements at CNR-SPIN. A slightly higher value, namely 19.35\,K was measured in Vienna for the unirradiated sample D, while only 19.05\,K was found for the irradiated sample B$^\mathrm{irrad}$. The small reduction ($\sim$0.3\,K)  agrees with the results on the Sm-1111 and Ba-122 systems \cite{Eis10c}.

\begin{figure} \centering \includegraphics[clip,width=0.5\textwidth]{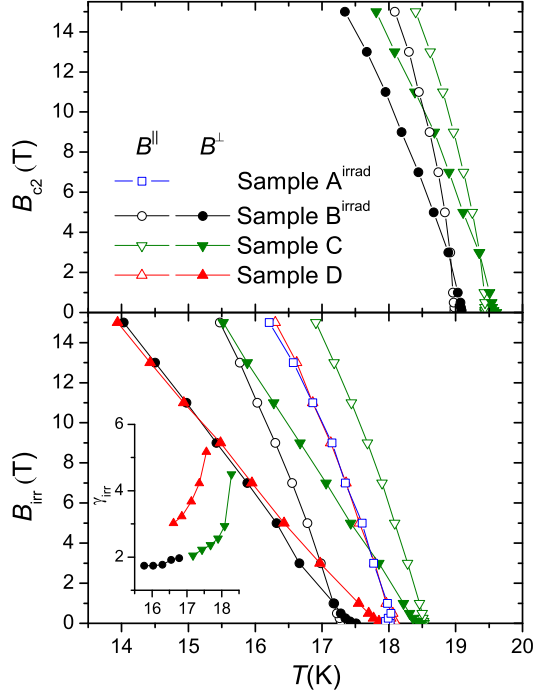}
\caption{Upper critical field (upper panel) and irreversibility (lower panel) lines for both main field orientations. Samples C and D were measured before, samples A$^\mathrm{irrad}$ and B$^\mathrm{irrad}$ after irradiation. The insert shows the anisotropy of the irreversibility field, $\gamma_\mathrm{irr}=\frac{B_\mathrm{irr}^\parallel}{B_\mathrm{irr}^\perp}$ near $T_\mathrm{c}$. Parallel ($\parallel$) and perpendicular ($\perp$) refer to the film plane.} \label{FigBirr}
\end{figure}

The temperature dependence of the upper critical field, $B_\mathrm{c2}(T)$, of samples B and C and the irreversibility lines, $B_\mathrm{irr}(T)$, of all samples are plotted in the upper and lower panel of Fig.~\ref{FigBirr}, respectively. $B_\mathrm{c2}(T)$ is extremely steep with a negative curvature for both main field orientations (for fields above 1\,T), as observed in Fe$_{1.05}$Se$_{0.11}$Te$_{0.89}$ \cite{Lei10} and  FeSe$_{0.4}$Te$_{0.6}$ \cite{Khi10} single crystals. This was ascribed to Pauli limitation, which is particularly important in the 11 phase \cite{Put10}, and would also explain the small effect of disorder on the upper critical field found in our films. However, a minor effect of disorder on $B_\mathrm{c2}$ was also found in other iron based compounds \cite{Eis09b,Eis09c}.  

The irreversibility lines are shifted to lower temperatures (by about 1.5\,K at 0\,T) and only slightly flatter than $B_\mathrm{c2}(T)$, which is a consequence of the somewhat increasing transition width, which doubles at 15\,T compared to 0\,T in the perpendicular field orientation (parallel and perpendicular refer to the film plane). This demonstrates sharp transitions even at high fields. In the parallel field orientation, the field induced broadening is even smaller. The only qualitative difference between $B_\mathrm{c2}(T)$ and $B_\mathrm{irr}(T)$ is a different curvature in the perpendicular configuration: The negative curvature in $B_\mathrm{c2}(T)$ is replaced by a positive curvature in $B_\mathrm{irr}(T)$ over the whole field range accessible in our experiments. It is interesting to note that $B_\mathrm{irr}(T)$ keeps the negative curvature of $B_\mathrm{c2}(T)$ in parallel orientation (cf. Fig.~\ref{FigBirr}), while it does not in perpendicular orientation. This indicates that only in the perpendicular configuration thermally activated depinning becomes significant at high temperatures. The importance of these fluctuations (e.g. the difference between  $B_\mathrm{c2}$ and $B_\mathrm{irr}$) seems small compared to the cuprates, where a positive curvature of the irreversibility line is well established in all compounds and all field orientations.

The anisotropy of the irreversibility fields increases with temperature (inset in Fig.~\ref{FigBirr}), as a consequence of the positive curvature of $B_\mathrm{irr}^\perp(T)$ near $T_\mathrm{c}$. At temperatures, where $B_\mathrm{irr}$ in parallel orientation reaches 15\,T, we find an anisotropy factor of 1.7 (at 15.5\,K), 2 (at 16.9\,K) and 3 (at 16.3\,K) in samples B, C, and D, respectively. This anisotropy of $B_\mathrm{irr}$ near $T_\mathrm{c}$ is smaller than that in a FeSe$_{0.5}$Te$_{0.5}$ single crystal (around 4) \cite{Kle10}, but a smaller anisotropy in films compared to bulk materials is not unusual. 

The irreversibility lines of the irradiated sample A$^\mathrm{irrad}$ and the pristine sample D are virtually identical in parallel orientation. The irreversibility line of the irradiated sample B$^\mathrm{irrad}$ is slightly steeper than $B_\mathrm{irr}(T)$ of sample D (unirradiated) in the perpendicular configuration, but $B_\mathrm{irr}(T)$ of the unirradiated sample C is steepest in this orientation. It can be concluded, that sample-to-sample variations are larger than the effect of the neutron induced defects.                   

\begin{figure} \centering \includegraphics[clip,width=0.5\textwidth]{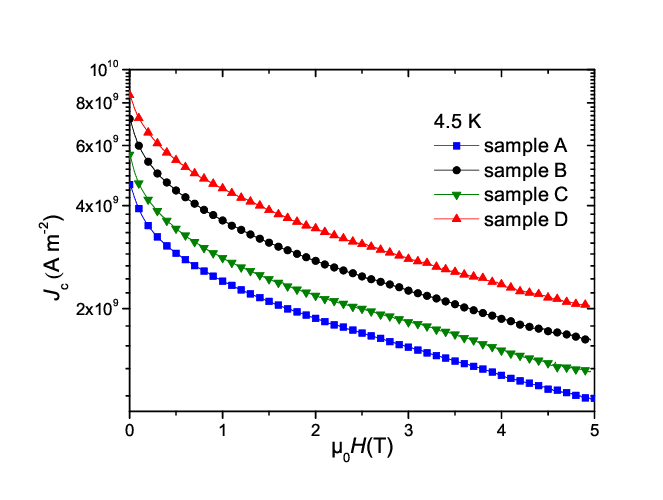}
\caption{Comparison of the critical current densities at 4.5\,K in all unirradiated samples. The field was oriented perpendicular to the films.} \label{FigJc5K}
\end{figure}

\begin{figure} \centering \includegraphics[clip,width=0.5\textwidth]{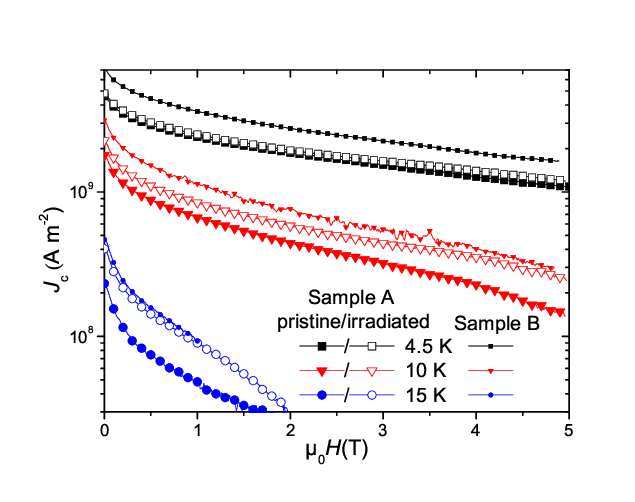}
\caption{Critical current densities in sample A prior to and after neutron irration. $J_\mathrm{c}$ of the best pristine sample with the same thickness (B, small symbols) is shown for comparison.} \label{FigJcirr}
\end{figure}

Figure~\ref{FigJc5K} presents the critical current densities of all samples (before irradiation) at 4.5\,K in perpendicular field orientation. They are higher (but more field dependent) than those in FeSe$_{0.4}$Te$_{0.6}$ single crystals \cite{Tae09,Liu10} ($\sim 10^{9}$\,A\,m$^{-2}$), which is rather expected for a high quality film. The critical currents in a stressed \cite{Mel10} and sulfur doped \cite{Han10} FeTe film as well as in a Co-doped Ba-122 film \cite{Iid10} were reported to be significantly lower ($<10^{9}$\,A\,m$^{-2}$) under comparable conditions. However, significantly higher values ($>10^{10}$\,A\,m$^{-2}$) in Ba-122 films were reported, too \cite{Tar10,Lee10,Kat10,Iid10b,Moh10}. All our samples have the same field dependence, but $J_\mathrm{c}$ in the best sample (D) is about 85\% higher than in the worst sample (A). The same field dependence in all samples is also found at 10\,K and 15\,K (e.g. Fig.~\ref{FigJcirr}). The temperature dependence of $J_\mathrm{c}$ is not the same in all samples, which excludes a scenario, in which the differences between the samples are only given by material inhomogeneities or defects influencing the geometry of current flow. The differences must be (at least partly) caused by differences in the pinning landscape.

The influence of neutron irradiation on the critical currents at 4.5\,K, 10\,K, and 15\,K is demonstrated in Fig.~\ref{FigJcirr}. It is rather small at 4.5\,K and becomes larger at higher temperatures (increase by a factor of about 2 at 15\,K). Although the positive effect of the irradiation is unambiguous, since the same sample was measured before and after irradiation, the changes remain within our sample-to-sample variations, because $J_\mathrm{c}$ in the irradiated sample A does not exceed $J_\mathrm{c}$ in the unirradiated sample B.  A stronger influence of neutron irradiation on the critical currents was found in sintered Sm-1111 \cite{Eis09b} and in Ba-122 \cite{Eis09c} single crystals, where both the relative enhancement as well as the achieved current densities were significantly larger. It is interesting to note that $J_\mathrm{c}$ in the 11 films exhibits a weak dependence on the field before the irradiation, whereas $J_\mathrm{c}$ is largely suppressed by an applied field of 1 T in unirradiated Sm-1111. The irradiation introduced defects, which effectively pin vortices in Sm-1111 and Ba-122, thus weakening the field dependence of $J_\mathrm{c}$. On the contrary, pinning is strong in the pristine 11 thin films and therefore not significantly enhanced by the irradiation. Indeed, at the highest temperature (15\,K), where $J_\mathrm{c}$ in the pristine sample is strongly suppressed by magnetic fields due to thermal depinning, a significant increase of $J_\mathrm{c}$ is obtained upon irradiation. 

The critical currents decreased by a factor of about 5 in sample B after the irradiation process. It seems extremely unlikely that the neutron collisions with the lattice atoms caused this suppression, due to the totally different behavior of sample A$^\mathrm{irrad}$. A reaction of the superconducting layer with air or moisture, which possibly contaminated the helium atmosphere in the quartz capsule (samples A and B were sealed into different capsules), is a plausible explanation for this degradation.         

\begin{figure} \centering \includegraphics[clip,width=0.5\textwidth]{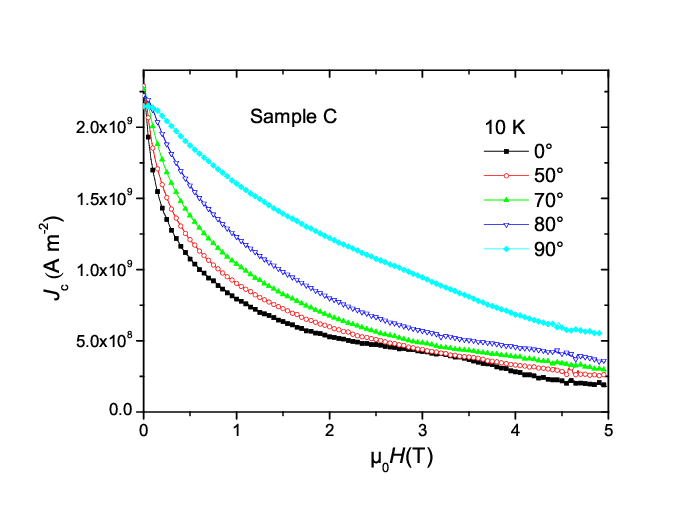}
\caption{Critical current densities in sample C at 10\,K and various angles between the tape normal and the applied magnetic field.} \label{FigJcang}
\end{figure}

The critical current densities in sample C at 10\,K are plotted in Fig.~\ref{FigJcang}. Data corresponding to different angles between the film normal and the magnetic field are shown. The critical currents remain roughly constant at low angles (not shown) and increase monotonously at higher angles. The opposite behavior is observed at very low magnetic fields, which is a consequence of the decreasing electric field (see Sec.~\ref{secexp}). 

\begin{figure} \centering \includegraphics[clip,width=0.5\textwidth]{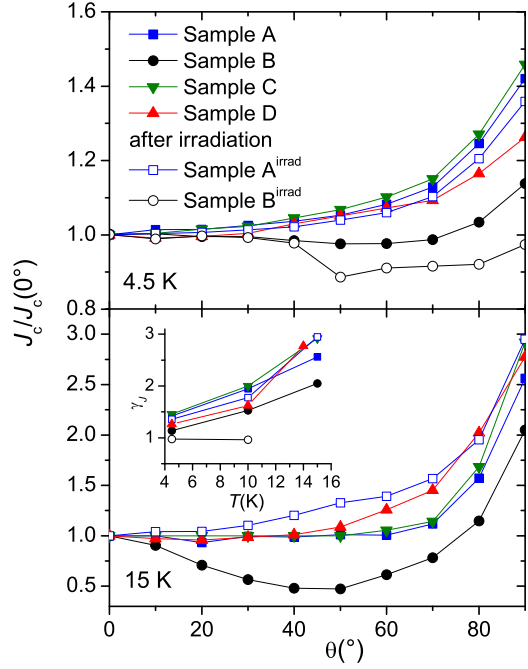}
\caption{Angular dependence of the critical currents at 1\,T. Data in the upper and lower panel refer to 4.5\,K and 15\,K, respectively. The $J_\mathrm{c}$-anisotropy increases with temperature as can be seen best in the inset.} \label{Figang}
\end{figure}

The sensitivity of the critical currents on the orientation of the magnetic field can be better seen by plotting the angular dependence of the critical current density at fixed magnetic field. $J_\mathrm{c}(\theta)$ at 1 T (normalized by $J_\mathrm{c}(0^\circ)$) in all samples is plotted in Fig.~\ref{Figang}. The data in the upper panel refer to 4.5\,K, in the lower panel to 15\,K. The critical currents in samples A, C, and D monotonically (within experimental accuracy) increase with angle and no indication of c-axis correlated defects is observed (which does not exclude their presence). The $J_\mathrm{c}$-anisotropy increases with temperature, by around one third from 4.5\,K to 10\,K, and is approximately doubled at 15\,K (inset in Fig.~\ref{Figang}). This does not necessarily mean that the effective electronic mass anisotropy, $\gamma$, changes, since a temperature dependence of $\gamma_J$ at fixed magnetic field is expected also for constant $\gamma$. Any anisotropy of the irreversibility line above one leads to a divergence of $\gamma_J$ at $B_\mathrm{irr}^\perp$, where $J_\mathrm{c}(0^\circ)$ per definition converges to zero. At zero applied field on the other hand, $\gamma_J$ has to converge to one, since the self field of the sample becomes dominant. It would be best to compare  $\gamma_J$ at the same reduced field, $B$/$B_\mathrm{irr}^\perp$, which is unfortunately not possible, since the irreversibility field at low temperatures is far outside our experimental window. It is therefore instructive to consider the field dependence of $\gamma_J$. No significant change is observed between 1\,T and 4\,T at 4.5\,K, while it slightly (about 15\%) increases at 10\,K. Since it is extremely unlikely that $B_\mathrm{irr}$ increases by more than a factor of four (between 10\,K and 4.5\,K) and $\gamma_J$(4.5\,K,4\,T) is smaller than $\gamma_J$(10\,K,1\,T), it is safe to assume that an increase in $\gamma_J$ would be observed also at fixed reduced magnetic field.  
 
One possible reason for the temperature dependence of $\gamma_J$ is a changing effective mass anisotropy $\gamma$, which was reported to drop from four to well below two between $T_\mathrm{c}$ and 0.95$T_\mathrm{c}$ \cite{Kle10} in single crystals and then slowly converges to one at low temperatures \cite{Lei10}. However, the increase of $\gamma_J$ by a factor of two between 4.5\,K and 15\,K is considerably more than expected from the single crystal data of $\gamma$ and correlated pinning centers (parallel to the film) seem to be a more likely explanation. Since they obviously pin very efficiently, they might compete better with the thermal energy at high temperatures than the uncorrelated pinning centers. This scenario is supported by the shape of $J_\mathrm{c}(\theta)$. The $ab$-peak is much sharper than expected from mass anisotropy, in particular for moderate mass anisotropies.  $J_\mathrm{c}$ is even virtually independent of the field orientation up to 60\textdegree\ in sample A at 15\,K and 1\,T (solid squares in the lower panel of Fig.~\ref{Figang}). Correlated pinning parallel to the ab-planes was also found in Ba-122 films \cite{Iid10,Iid10b} by an anisotropic scaling analysis\cite{Bla92}.
 
The observed sample-to-sample variations in our films are not pronounced, except for sample B, which behaves qualitatively differently. A broad and hard-to-resolve c-axis peak is observed in the angular dependence of $J_\mathrm{c}$ at 4.5\,K in this sample and becomes quite pronounced at 15\,K. We cannot exclude that the peak is an artifact of the measurement at 4.5\,K (see Sec.~\ref{secexp}), but the decrease in  $J_\mathrm{c}$ between 0\textdegree\ and 40\textdegree\ is much faster than expected from the variation of the electric field, even in the limiting case of ohmic behavior. The peak is a clear indication for c-axis correlated pinning. Also Ba-122 films with \cite{Lee10,Tar10} and without \cite{Iid10,Iid10b} a maximum of $J_\mathrm{c}$ in perpendicular orientation were reported.   
   
A decrease in $\gamma_J$ is observed in both samples at 4.5\,K and 10\,K after irradiation, only at 15\,K the $J_\mathrm{c}$-anisotropy of sample A increases. (The signal of sample B$^\mathrm{irrad}$ was too small at 15\,K to be measured reliably.) The decrease at low temperatures is not unexpected, since disorder generally tends to smear anisotropic properties. The increase at elevated temperatures is difficult to understand. The decrease in transition temperature might contribute to this effect, since it might decrease the irreversibility field at 15\,K by the shift in transition temperature (see above). Since $\gamma_J$ diverges at $B_\mathrm{irr}^\perp$, a small change in $B_\mathrm{irr}^\perp$ can result in a large increase in $\gamma_J$. The irreversibility field in perpendicular orientation is around 10.7\,T, if derived by a 10\% resistive criterion (cf. Fig.~\ref{FigBirr}), which seems to exclude this effect at 1\,T. However, the temperature, at which the resistivity at 1\,T becomes zero within experimental accuracy, is about 16\,K in this sample (after irradiation); thus the small reduction in $T_\mathrm{c}$ could contribute significantly to the  unconventional increase of $\gamma_J$ at 15\,K. 

\section{Conclusions}
Critical current densities well above $10^9$\,A\,m$^{-2}$ were found at 4.5\,K up to 5\,T. The anisotropy of $J_\mathrm{c}$ is small at low temperatures and the currents are highest, if the field is applied parallel to the film surface. The anisotropy increases at higher temperatures and a pronounced c-axis peak appears in one sample, which indicates correlated pinning centers parallel to the film normal. Correlated pinning parallel to the film plane (parallel to the crystallographic ab-planes) was found in all samples and could be intrinsic in nature. The presence or absence of the c-axis peak on the other hand, is a consequence of different pinning centres and indicates their sensitivity on the preparation conditions.
Thermally activated depinning seems rather unimportant in these films, but likely causes the positive curvature of the irreversibility line in the perpendicular field orientation. 
The effect of disorder, which was introduced by neutron irradiation, on the upper critical field, the irreversibility field and the critical currents is small and comparable to typical sample-to-sample variations.    

\begin{acknowledgments}
The work at the Atominstitut was supported by the Austrian Science Fund (FWF):22837.
\end{acknowledgments}

%\bibliographystyle{michl}
%\bibliography{ref} % ref.bib is the name of our database

\end{document}